# INSTITUTIONALIZATION IN EFFICIENT MARKETS: THE CASE OF PRICE BUBBLES

SHEEN S. LEVINE
Singapore Management University
50 Stamford Road, Singapore

EDWARD J. ZAJAC
Northwestern University

## ABSTRACT

We seek to deepen understanding of the micro-foundations of institutionalization while contributing to a sociological theory of markets by investigating the puzzle of price bubbles in financial markets. We find that such markets, despite textbook conditions of high efficiency – perfect information, atomistic agents, no uncertainty – quickly develop patterns consistent with institutionalization processes.

## INTRODUCTION

In the last three decades years, the "new" institutional theory rose to prominence, becoming a cornerstone on the relationship between organizations and their environment. Its tenets – the organizational pursuit of legitimacy and status (Meyer & Rowan, 1977), the spread of believes, norms and practices (DiMaggio & Powell, 1983), and the institutionalization of processes and structures into a rule-like status (Zucker, 1977) – have become underlying assumptions of organizational theory at large. Institutional theory had wide ranging impact and was used to explain a variety of important phenomena including organizational structure, culture, change and response to external pressures, adoption of practices, and the impact of institutions on individuals in organizations.

*Institutionalization.* The importance of institutionalization, as both a process and a property variable, was noted early in the development of the field (Zucker, 1977). However, despite the proliferation of institutional theory, the micro-foundations of institutionalization have received scant empirical attention: "the experimental social psychological work of Zucker (1977) and some of her subsequent work focusing on how context differences lead to multiple social orders offer insights into institutional variation at more local levels. This line of research is the most underdeveloped in institutional theory" (Lounsbury, 1997:468). Because this line of research has seen little development, theory has necessarily resorted to presumptions about the mechanisms of institutionalization, leading to a structurally deterministic theory. While recent years have seen rise of research on change at the macro level, there is paucity of evidence on the origins of institutionalization. Despite Zucker's assertion that "macro-level and micro-level are inextricably intertwined" (1977:728), multi-level assessments remain elusive. Our work here builds on earlier efforts to alleviate this paucity by directly examining micro-processes of institutionalization.

*Markets.* The emergence of the new economic sociology has brought renewed interest in markets as a phenomenon that deserves sociological attention (Baker, Faulkner, & Fisher, 1998; Levine, Apfelbaum, Bernard, Bartelt, Zajac, & Stark, 2014). Although financial markets, with their vast economic and social impact, carry particular importance, the lenses of institutional theory have not been turned to such markets frequently, with few notable exceptions (Abolafia & Kilduff, 1988; MacKenzie & Millo, 2003b; Zajac & Westphal, 2004a).

The micro-foundations of institutionalization in markets remain conjectural to date. We seek to deepen and extend Zucker's original insight on the micro-processes of institutionalization while contributing to the sociological understanding of markets, by using a high-efficiency financial market as our empirical context. One of the difficulties in treating micro-level action is the garnering of fine-grained individual data while observing macro-level outcomes (cf. Lucas & Lovaglia, 2006). Hence, in this study we utilize a well-established experimental methodology from behavioral economics to observe and measure such micro processes. It also allows us to rule out purely individual-level explanations for inefficiency that are steeped in economic psychology or behavioral economics.

## THE PUZZLE OF PRICE BUBBLES

Contemporary economic theory recognizes that markets sometimes develop price bubbles, i.e., they exhibit "trade in high volume at prices that are considerably at variance from intrinsic value" (King, Smith, Williams, & van Boening, 1993:183). Cases such as the stock market crash of 1929 demonstrate the enormous effect of bubbles on individuals, firms, markets and even nations, and explain the interest they draw from economists as well as the public. While important in their consequences, the causes of bubbles are not well understood and there is no widely accepted theory to explain their occurrence. The existence of market bubbles seems at odds with common assumptions regarding the efficiency of financial markets. Even more puzzling is the finding that bubbles occur not only in real-world markets, with their inherent noise and uncertainty, but also in highly predictable experimental markets (Smith, Suchanek, & Williams, 1988). As Nobel Laureate Vernon Smith and associates conceded, "controlled laboratory markets price bubbles are something of an enigma" (Smith, van Boening, & Wellford, 2000:568).

In a typical study, participants engage in trading of assets that are defined to have a finite lifespan and a known distribution of dividends. Uncertainty is eliminated and participants should be able to calculate the *intrinsic value* of the assets simply by examining the expected stream of dividends. The market epitomizes economic efficiency, because in addition to complete information each individual can simultaneously buy and a sell (as common in stock markets). Communication is strictly limited to the posting of bid and ask offers, so collusion is unlikely and efficient prices should prevail. Nevertheless, price bubbles have been observed repeatedly in experimental markets, even with sophisticated participants such as business students, managers, and professional traders. Such bubbles have proven robust to a variety of conditions, including short-selling, margin buying, equal portfolio endowment, brokerage fees, the presence of informed insiders, dividend certainty, constant value, and limit price change rule (King et al., 1993; Porter & Smith, 2003).

*Bounded rationality.* While a complete explanation of experimental bubbles is still in the making, some implied that the phenomena is the result of bounded rationality, as it has been shown that bubbles abated when participants traded repeatedly within the same group (Dufwenberg, Lindqvist, & Moore, 2005; King et al., 1993; van Boening, Williams, & LaMaster, 1993). It may be argued that bounded rationality gives rise to bubbles in the short term as people may have cognitive difficulties in applying a theoretical pricing model. Arguably, initial mispricing may be decreasing due to individual learning processes, which lead to improved pricing in subsequent periods and an overall abatement of bubbles.

*Greater Fool explanation.* Prevalent among practitioners but present also in the academic literature (Smith et al., 1988:1148), the Greater Fool explanation posits that bubbles are fueled by speculators who *knowingly* purchase overpriced assets in the hope that they can later sell those assets even more dearly to gullible investors, i.e. "greater fools". The psychological literature provides support through evidence that people tend to be overconfident about their own skills (Svenson, 1981).

*Institutionalization.* It is also possible that bubbles stem not from psychological or otherwise individually-based biases, but from institutionalization in which "correct" interpretations of facts and legitimate reactions become taken-for-granted, as was shown in previous experiments (Zucker, 1977) and financial markets studies (Abolafia & Kilduff, 1988; Levine et al., 2014; MacKenzie & Millo, 2003a; Westphal & Zajac, 2001; Zajac & Westphal, 2004b). Institutionalization can lead to coordinated action through the internalization of beliefs and interpretations of facts by each individual, even without explicit discussion of coordination and even when the practices are inefficient or plain wrong (Meyer & Scott, 1992).

The three explanations should exhibits marked empirical differences in measurements of participants' pricing skills and overconfidence bias ex-ante and in the correlation of price discrepancy, i.e., the distance between the market price of an asset and the intrinsic value of that asset. Pricing discrepancies can be decomposed into *dispersion* and *common* components, which measure deviation from a common pricing method and correlation between discrepancies, respectively. If bubbles are caused by bounded rationality, then one can expect indication of inadequate pricing skills ex-ante and high dispersion component, where market prices fall randomly above or below the intrinsic values. If the Greater Fool explanation holds, one can expect evidence of widespread overconfidence bias. If bubbles are caused by institutionalization, the common discrepancy component should be high and price discrepancies should be correlated both among individuals and over time, as institutionalization spreads and settles.

## METHOD

Following in the footsteps of institutional researchers (e.g. Elsbach, 1994; Lucas, 2003; Lucas & Lovaglia, 2006; Zucker, 1977) and numerous precedents in behavioral economics, we chose an experimental approach. We constructed a laboratory double auction market, which is known to possess characteristics of extreme economic efficiency or competitiveness (Holt, 1995). The experimental market was programmed and conducted in z-Tree (Fischbacher, 2007) and based on designs published in the economics literature (Dufwenberg et al., 2005; Smith et al., 1988). Taking a careful and rigorous approach, we created a conservative test of institutionalization by using a synthetic, high efficiency market coupled with little social interaction: participants could not communicate directly, had no shared history, and were guaranteed complete anonymity to avert any future consequences. Further, we did not use deception to initiate institutionalization (cf. Zucker, 1977). Rather, we expected it to appear *in vivo*, even though the experimental conditions eliminated conditions discussed in prior research in economics and sociology, such as market power, uncertainty, repeated interaction, identity and culture.

We recruited 62 undergraduate students with no prior experience in such experiments for ten separate experimental sessions. At the experimental laboratory, participants received description of the experiment and information necessary for the calculation of asset values. After reviewing the information, the participants received a Price Questionnaire that probed their knowledge of a standard asset-pricing model. Then, the participants were asked to complete an

Assessment Questionnaire, which included questions designed to assess overconfidence. After completing the pre-trade questionnaires, the participants moved to a behavioral laboratory, where they sat in separate cubicles in front of networked personal computers and began trading. They were free to enter their minimum selling prices (ask) and/or their maximum offers to buy (bid). The participants knew that their earnings would be paid to them in cash at the end of the experiment.

Each of the ten experimental sessions consisted of six participants who traded for 10 periods lasting 120 seconds each. At the end of each period, a dividend of 20 cents per share was payable with a probability of 0.5. A summary screen appeared at the end of each period and presented individualized trading and divided results. At the conclusion of the trading periods, the participants received a $5 show-up fee and their earnings in cash. On average, payments were $13.38 (s.d.=2.87; range=$5.30-$18.30).

## RESULTS

Similar to prior work, we observed bubbles in most of the experimental sessions. Figure 1 illustrates the declared, intrinsic, and actual prices over trading periods, and shows how prices occasionally rose above the maximum present value (the sum of the highest possible dividends).

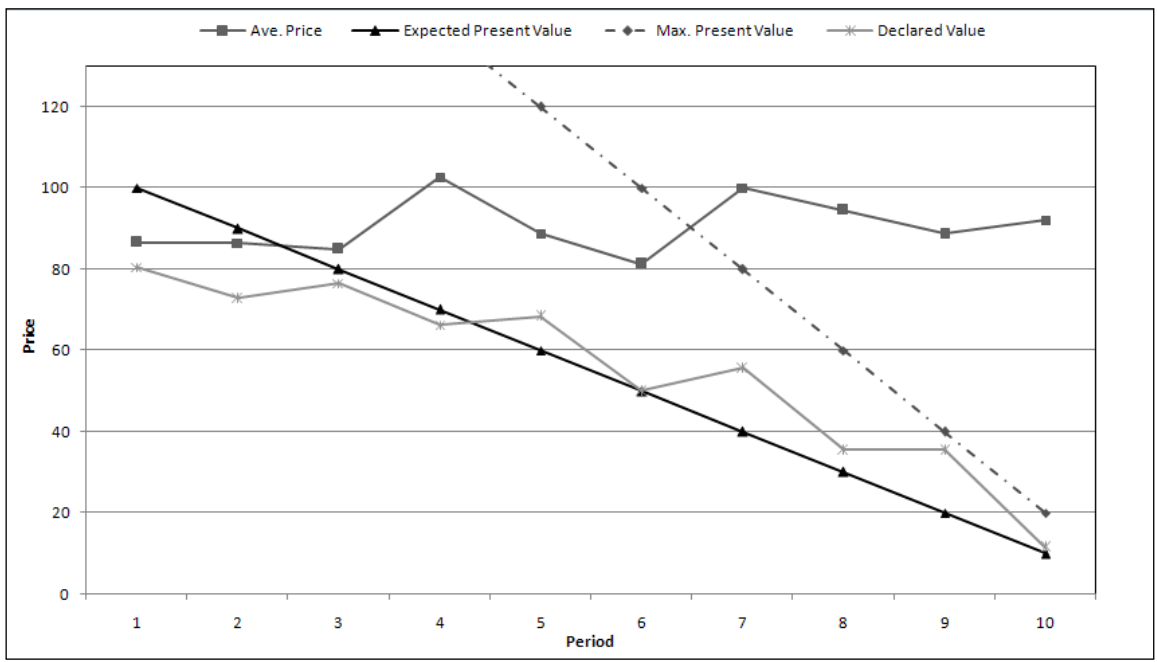

**Figure 1:** Illustration of declared, intrinsic, and actual prices over trading periods
Legend (left to right): average trading price; average present value (intrinsic value); maximum present value possible (intrinsic value); average declared value in the Price Questionnaire.

*Ex-ante pricing skills*. We found no indication that bubbles were caused by lack of knowledge. Quite the contrary: participants had better understanding of the theoretical pricing model *ex-ante*, but – astonishingly – seem to have abandoned that understanding during trading. We used Haessel's $R^2$ (Haessel, 1978) to measure fit between the responses to the Price Questionnaire and intrinsic values and between the trading prices and intrinsic values. A comparison of average prices declared in the Price Questionnaire with those obtained in actual trading revealed that the prices declared *ex-ante* were better fit to intrinsic values in nine out of 10 sessions. Prices in trading also had higher normalized average price deviation in nine out of 10 sessions and wider price amplitude in all of the sessions.

*Overconfidence bias.* We found no indication of overconfidence bias. Participants generally viewed their own price assessment to be as precise as the others'. Measures of Cronbach's alpha (Cronbach, 1951) show high reliability for each group of questionnaire items.

*Decomposition of price discrepancy.* Importantly, we found that discrepancies between market prices and intrinsic values were increasingly correlated among participants over time. Decomposition of the average discrepancy between a given market price and the matching intrinsic value (Hommes, Sonnemans, Tuinstra, & van de Velden, 2005) shows that the *common* component played a major role in the discrepancies between market price and intrinsic values. It accounts for most of the variance in all sessions but one. Moreover, as Figure 2 demonstrates, the common component tends to *increase* during trading, showing that errors became more correlated as trading progressed.

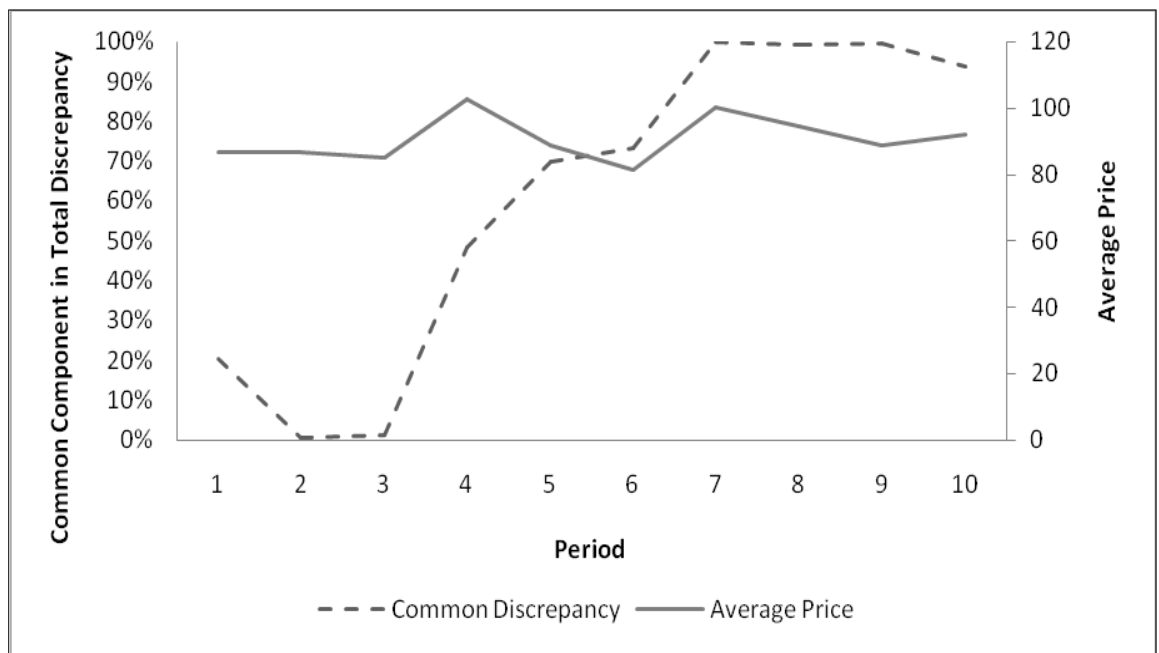

**Figure 2:** Illustration of common component in price discrepancy over trading periods

## DISCUSSION AND IMPLICATIONS

The results suggest that price bubbles were not simply the result of individual cognitive bias. We find that financial markets, even with textbook conditions of high efficiency, quickly develop patterns of behavior that are consistent with our predictions of institutionalization. We show that purely psychological explanations, such as bounded rationality and overconfidence, are insufficient to account for these patterns.

Our findings problematize a tenet of neo-classical economics – that markets can be assumed to be efficient when certain characteristics are present: atomicity, product homogeneity, perfect information, equal access to technology and resources, free entry, and no regulation

(Hirschman, 1982). This presumption is central in public policy and is used to both justify regulation, as in anti-trust action, and advocate less government intervention, as in privatization. Even in sociological thought, which is generally indifferent about departure from market efficiency, it presumed that such departure will occur only under certain circumstances, such as the embeddedness of social and economic relations (which violates the atomicity assumption) (Uzzi, 1997). Our results, however, show that markets, even in an ideal state of presumed efficiency, may be more institutionalized than commonly thought. In fact, the findings here imply that institutionalization can happen even in extreme case of what Granovetter (1985) called *undersocialized* conditions.

      We do not doubt that markets can sometimes lead to efficient allocation of resources. However, if institutionalization appears so rapidly and settles so profoundly in a synthetic market designed for high efficiency, it seems reasonable to expect that even greater institutionalization will occur in real-world markets, financial or others, with their inherent uncertainty, lower efficiency, and direct communication between market participants. Thus, markets may be realms of coordinated action, much like organizations, further blurring the distinction between the two social structures.

      It is quite possible that institutions sometimes require long time or intensive interaction to develop, but our results provide additional support to Zucker's finding that, given the right conditions, institutionalization can happen quickly and even with little social interaction. This suggests a clear distinction between institutions and culture: while the two may be intertwined, institutions are not always the proverbial stone shaped by water over time.

      A fruitful path would be to investigate how social interaction leads to institutionalization in markets and how it can be reversed. Recently, it was shown that in markets for cultural products, small initial differences can lead to dramatically different outcomes (Salganik, Dodds, & Watts, 2006). Additional work will be necessary to establish the conditions that induce institutionalization in markets, its magnitude, and how it might be reversed.